# STATISTICAL CONTEXT-DEPENDENT UNITS BOUNDARY CORRECTION FOR CORPUS-BASED UNIT-SELECTION TEXT-TO-SPEECH

Claudio Zito[1], Fabio Tesser[2], Mauro Nicolao[2,3], Piero Cosi[2]

[1]Dipartimento di Informatica, Università di Pisa, Italia
[2]Istituto di Scienze e Tecnologie della Cognizione, Consiglio Nazionale di Ricerca, Italia
[3]Speech and Hearing Research Group, University of Sheffield, United Kingdom
*c.zito@studenti.unipi.it, fabio.tesser@gmail.com, mauro.nicolao@gmail.com, piero.cosi@pd.istc.cnr.it*

## 1. ABSTRACT

In this study, we present an innovative technique for speaker adaptation in order to improve the accuracy of segmentation with application to unit-selection Text-To-Speech (TTS) systems. Unlike conventional techniques for speaker adaptation, which attempt to improve the accuracy of the segmentation using acoustic models that are more robust in the face of the speaker's characteristics, we aim to use only context dependent characteristics extrapolated with linguistic analysis techniques. In simple terms, we use the intuitive idea that context dependent information is tightly correlated with the related acoustic waveform. We propose a statistical model, which predicts correcting values to reduce the systematic error produced by a state-of-the-art Hidden Markov Model (HMM) based speech segmentation. In other words, we can predict how HMM-based Automatic Speech Recognition (ASR) systems interpret the waveform signal determining the systematic error in different contextual scenarios. Our approach consists of two phases: (1) identifying context-dependent phonetic unit classes (for instance, the class which identifies vowels as being the nucleus of monosyllabic words); and (2) building a regression model that associates the mean error value made by the ASR during the segmentation of a single speaker corpus to each class. The success of the approach is evaluated by comparing the corrected boundaries of units and the state-of-the-art HHM segmentation against a reference alignment, which is supposed to be the optimal solution. The results of this study show that the context-dependent correction of units' boundaries has a positive influence on the forced alignment, especially when the misinterpretation of the phone is driven by acoustic properties linked to the speaker's phonetic characteristics. In conclusion, our work supplies a first analysis of a model sensitive to speaker-dependent characteristics, robust to defective and noisy information, and a very simple implementation which could be utilized as an alternative to either more expensive speaker-adaptation systems or of numerous manual correction sessions.

## 2. INTRODUCTION

The latest generation of speech synthesis techniques has recently increased the quality of Text-To-Speech (TTS) systems as regarding the naturalness and the intelligibility of the voice. Such systems are often referred in the literature as corpus-based TTS.

The unit-selection method (Hunt & Black, 1996) is one of the technologies for corpus-based TTS that accumulates human speech (natural speech) in a database of reusable units, and generates synthesized speech by properly concatenating the units. Considering all the possible techniques for corpus-based TTS systems, the unit selection methods are the ones most sensitive to the dimension and the quality of the speech corpus because the algorithms and the databases are determined by a statistical approach based on a large-scale speech corpus. During database creation, each recorded utterance is segmented into some or all of the following: individual phones, diphones, half-phones, syllables, morphemes, words, phrases, and sentences.

The speech segmentation phase plays a crucial role for the naturalness of the unit-selection voice. In fact, speech segmentation defines the acoustic unit boundaries, which divides up sounds into the audio files that the speech corpus contains. An inaccurate alignment of the corpus produces non-complete acoustic samples or marred by the presence of sounds belonging to neighboring units, degrading the intelligibility and naturalness of the system output.

The present study proposes a statistical method based on regression trees (Breiman et al., 1984) to improve the naturalness and the intelligibility of unit-selection TTS systems. A system was built which automatically predicts correcting values in order to improve the accuracy of the speech segmentation of speaker-dependent corpus. Our approach aims to achieve a more accurate segmentation that can be summarized in two phases as follows: (1) identifying context-dependent phonetic unit classes, for instance the class which identifies vowels as being the nucleus of a monosyllabic word; and (2) building a regression model that associates with each identified class the mean error value made by the ASR during the segmentation. The main idea is to identify and reduce as much as possible the systematic error generated by an ASR during the speech segmentation phase of speaker-dependent corpora.

Typically, the division into segments is done using a specially modified general-purpose Automatic Speech Recognizer (ASR), based on Hidden Markov Models (HMMs) (Rabiner, 1989), set to a forced alignment mode with some manual correction afterwards, using visual representations such as the waveform and the spectrogram. The manual correction phase is extremely expensive since it requires a large amount of man-hours by a human operator. Furthermore, although such a system is usually speaker-dependent because the entire corpus is a recording of a single speaker, or very few speakers, general-purpose ASR does not exploit this important characteristic to improve the segmentation, making necessary a further speaker-adaptation phase. An index of the units in the speech database is then created based on the segmentation and acoustic parameters like the fundamental frequency (pitch), duration, position in the syllable, and neighboring phones. At runtime, the desired target utterance is created by determining the best chain of candidate units from the database. This process is typically achieved using a specially weighted decision tree.

In order to design this model we needed to define a database of utterances, which pledges a good coverage of the phonemes of the Italian language in different contexts and, for each speaker, a reference forced alignment, which is as accurate as possible. Subsequently, we built the speech corpus "ad hoc", as well as, an accurate segmentation which hereafter is referred to as the *reference transcription*. In addition, we built a prototype voice for a unit-selection TTS with the same speech corpus using the procedure suggested by the authors of FESTIVAL (Black et al., 2007). During this procedure, we did not use the reference alignment, but the speech segmentation was automatically re-computed with a different HMM-based ASR and then corrected by the regression model we introduce in this work, called Context-Dependent Units Boundary Correction Model. Note that in this con-text, the word "boundary" stands for the marker, which identifies the passage between two neighboring phonemes on the time line of the audio file.

Our model addresses a standard regression-type problem where the goal is attempting to predict the values of a dependent continuous variable from one or more independent continuous variables. The general approach is to derive predictions from a few simple if-then conditions where the threshold and the leaf values are "learned" with a supervised learning paradigm. Our model was trained "off-line" with respect to the construction of the prototype voice, using only a part of the utterances of our corpus and relating the automatic segmentation with the reference transcription. We evaluated the model from a statistical point of view in order to identify a measurement of the reduction of the error between the automatic speech segmentation of unknown utterances and their respective reference alignments, which is supposed to identify the optimal solution.

In this paper, the terms segmentation, transcription, forced alignment (or simply alignment) are used synonymously.

## 3. THE CORPUS

LOQUENDO supplied the database for research purposes as part of a project in collaboration with the CNR-ISTC of Padova, with the aim of experimenting with methodologies for speech signal analysis and synthesis. The corpus was recorded by an Italian male speaker with normal intonation in a silent room (semi-anechoic), then it was utilized to build the first prototype of an Italian voice with unit-selection synthesis for the open-source platform called FESTIVAL (Black et al., 2007).

For recording the corpus, we utilized a microphone Sennheiser MKH 40 P48, connected either to a data acquisition board of a Personal Computer or to a channel (right) of a DAT Sony DTC 1000 ES, and an Electro-glottograph connected to only a single channel (left) of the DAT. All the recordings were sampled at 44.1 kHz and then treated, first manually to remove artificial noise and then with automatic noise-reduction models based on spectral subtraction.

The corpus consists of 500 utterances, each of which is around 10-15 words in length, extracted from a collection of national newspaper articles. These utterances were selected so as to provide a suitable coverage of all the phonemes of the Italian language.

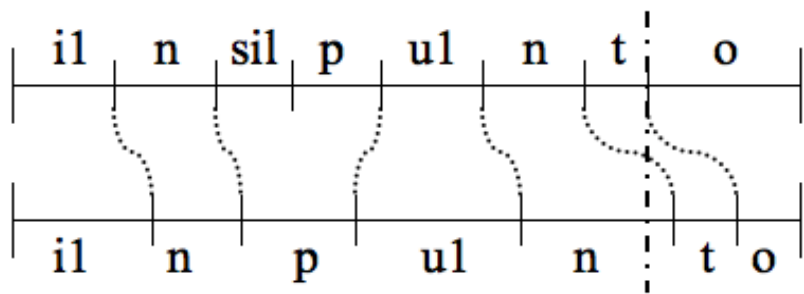

Figure 1: Simple example of the two classes. It shows how the reference transcription (up) and the sphinx transcription (down) for the same (piece of) utterance were compared phoneme by phoneme. It is possible to notice a recognition error for the phonemes /t/ and /sil/. The other cases belong to the position error class. The phonemes /i1/ and /u1/ identify the stressed vowels.

## 4. METHODOLOGY

### 4.1. Forced alignment of the corpus

In order to develop our regression model, we needed to perform the forced alignment of the corpus with two ASRs, which generate two alignments with different accuracy. The reference alignment identifies the reliable position of the marker for each boundary on the time line of the audio file and the second one identifies the state-of-the-art HMM-based speech segmentation.

The reference alignment used for the correction was supplied by LOQUENDO with the corpus described in the Section 3. LOQUENDO performed the speech segmentation with an HMM-based ASR system using a loop-forward algorithm (Rabiner, 1989). The LOQUENDO's speech segmentation algorithm provides a first segmentation by using a set of speaker-independent modules in order to recognize all the phonemes in the speech corpus. Then, these modules are trained onto the speaker acoustic characteristics in order to achieve a more accurate segmentation.

The second forced alignment was obtained with a less accurate procedure. We decided to use a simpler system in order to limit the numbers of tuning parameters and obtain easily a rough segmentation of the corpus. Therefore, we decided to use SPHINX-2 (CMU Sphinx, 2007) (Lee et al., 1990) with semi-continuous 5-state HMM-based module. Note that, when the number of states into the HMMs used for phonetic unit recognition is defined a priori, this affects the minimum required length of a phone, in order for it to be recognized by the system. To avoid this, we added specific transitions into the model to jump forward to non-neighboring states in order to achieve a better performance also with short-duration phones (Rabiner, 1989). Hereafter we refer to this alignment as *sphinx transcription*.

### 4.2. Error evaluation

In order to evaluate the performance of SPHINX-2 and collect the data to build the regression model, we looked at the differences between the sphinx transcription and the reference one. In simple terms, we compared the two transcriptions phoneme by phoneme and, for each marker, which defined the phoneme boundary, we simply computed the errors of the automatic procedure as the deviation from the correspondent reference alignment. Two different classes of error were defined as follows:

1. **Position errors**. The phoneme is correctly identified, but there is not an absolute coincidence between the corresponding boundary markers. This class of error identifies, for each category of phonemes, the trend of SPHINX to set in advance or postpone systematically the right-side marker of the given class of phones. At the end of the learning phase, this error defines an average of measurement of such a displacement;
2. **Recognition errors**. The phoneme is not correctly identified. It could happen because the phoneme is set out of its existence area or the two phonemes sequences do not coincide.

Figure 1 shows the comparison of the reference transcription (up) and the sphinx transcription (down) for the same (piece of) utterance. In the example, "in punto" is the Italian translation for the English expression "o'clock". The errors belonging to the recognition errors class were not evaluated in order to build the regression model, because such errors have no information about the systematic error made by the system. Regarding the position errors, they were collected as the distance in seconds between the position of the relative marker into the reference alignment and into the one to be corrected. A position error was considered significant only if it was greater than a hundredth of second.

More formally, let Err be the signal, which defines the behavior of the systematic error between the two forced alignments. And let L(ph) be a function defined as L:CDPH→TIME, where CDPH represents the context-dependent phones population, that for each phoneme gives as an output the desired position (LOQUENDO) in seconds of the marker for the right-side boundary in the acoustic file under analysis. In a similar way, let Sp(ph) be the function defined as Sp:CDPH→TIME for the alignment to be corrected (SPHINX). Hence the function Err could be simply computed as follow:

(1)
$$\mathbf{Err(ph) = L(ph) - Sp(ph)}$$

It is worth nothing that the error is not a proper distance measurement because it is not computed as an absolute value between the two alignments. The reason for this is that we need to cope also with the direction on which applying the predicted correction value, therefore the sign of the mean error is useful to identify the trend of the system to put in advance or postpone systematically the position of the marker for a given class of phonemes. Considering the empiric form of the treatment and the definition of systematical error this formulation turns out to be reliable. In fact, when the systematical error is present into the alignment, the collected errors will be significant and will have the same sign (direction). On the other hand, when the model is not going to recognize a given class of phonemes, the collected errors will be not significant and the sign will come out to be completely insignificant.

The context-dependent information is extracted in an automatic way thanks to the Heterogeneous Relation Graph (HRG) structure available on Festival, which is a formalism for representing linguistic information (Taylor et al., 2001). In the HRG formalism linguistic objects such as words, syllables and phonemes are represented by objects termed *linguistic items*. These items exist in relation structures, which specify the relationship between one another. A heterogeneous relation graph contains all the relations and items for an utterance.

Examples of the context dependent information we used in our experiments are: the current syllable's stress, the previous phonemes' class, the next phonemes' class, the number of stressed syllables in the utterance before the current phoneme and so forth.

The set called CDPH is composed of 71-feature vectors that describe the relation structure that affects the prosody of the given phoneme. The phoneme characterized in such a way comes out to give more information than the simple triphone or pentaphone (three or five neighboring phones) and is referred to as a *senone* in the literature (CMU, 2007).

## 5. BUILDING A REGRESSION MODEL

The regression model was built using the open source software tool called *wagon* by EDINBURGH SPEECH TOOLS LIBRARY (Black et al., 2003). This software was designed to build CART (Breiman et al., 1984) from a set of features extracted from a given dataset. Figure 2 shows in a graphic format the steps required to build a regression tree. Before explaining the details of the building process, we would like to draw the dataset as we used to training and testing our regression model.

| DATASET | | TRANING SET | | TEST SET | |
|---|---|---|---|---|---|
| Name | Units | Units | % | Units | % |
| LOQ | 25,360 | 22,428 | 88.4 | 2,932 | 11.6 |
| SPHINX | 26,815 | 23,739 | 88.6 | 3,076 | 11.4 |

Table 1: Decomposition of the corpus in acoustic units for the two forced alignments. In the table the number and the percentage of units present in the different datasets are shown.

In order to reuse the front-end with the linguistic modules defined in the diphone-based prototype voice in Italian proposed by the ISTC-CNR of Padova (Cosi et al., 2000) for Festival, we used a different phonetic mapping with respect to LOQUENDO. As the two different phonetic mappings use a different number of phonemes, this introduced differences in the transcription. In Table 1, the label LOQ refers to the dataset composed by the units extracted from the reference alignment, while the label SPHINX refers to the units from the sphinx alignment. 90% of the utterances present in the dataset were utilized for training the regression model, while the rest 10% were utilized for evaluating the performance on unknown utterances. It is worth noting from Table 1 that the percentage of the phonetic units present into the training set and the test set is well balanced for both the alignments.

| CARTree | RMSE | Corr | Mean Error | Mean Error (abs) |
|---|---|---|---|---|
| cor.S5.tree | 0.042 | 0.409 | 0.015 | 0.039 |
| cor.S10.tree | 0.050 | 0.341 | 0.016 | 0.042 |
| cor.S25.tree | 0.040 | 0.444 | 0.014 | 0.038 |
| cor.S100.tree | 0.042 | 0.355 | 0.015 | 0.039 |

Table 2: The Root Mean Squared Error (RMSE), correlation, mean error and mean error in absolute value for each mode. All the above values are in seconds.

The features data shown in Figure 2 consists of a set of training examples. Each example is a pair defined by a context-dependent features vector (senone) and a desired output value (systematic error). The features data were extracted directly from the dataset with a 3-step procedure consisting of the following phases:

1. **Pre-processing** enables the removal of the inconsistencies between the two alignments (i.e. recognition errors) and build an unique transcription of the training set. This procedure discards about 2.5% of units, reducing the training set to 23,151 phonemes;
2. **Features extraction** provides the linguistic analysis of each phoneme present into the unique transcription;
3. **Statistics** enables the labeling of each identified senone with the respective systematic error.

We built four models with different cluster size boundaries (5, 10, 25, 100). Table 2 shows the Root Mean Squared Error (RMSE), correlation, mean error and mean error in ab-solute value for each model. The nomenclature of the models is kept consistent with the one specified in Festival. For example, "cor.S5.tree" specifies a regression tree with maximum five candidates in each cluster.

Table 2 shows correlation values that are not larger than 0.444. These values should not be considered unreliable. In fact, the regression trees have to model a non-linear function, which represents the relationship between each senone (context-dependent phoneme) and the relative systematic error in seconds made by SPHINX-2. For those senones, which identify the most common context situations, the ASR had a more efficient training (with more data) and hence it performs a more accurate segmentation form, which it might not be possible to determine a systematic error. The results discussed in Section 6 show that such situations are usually determined when the collected errors are shorter the 0.01 seconds and have an arbitrary distribution.

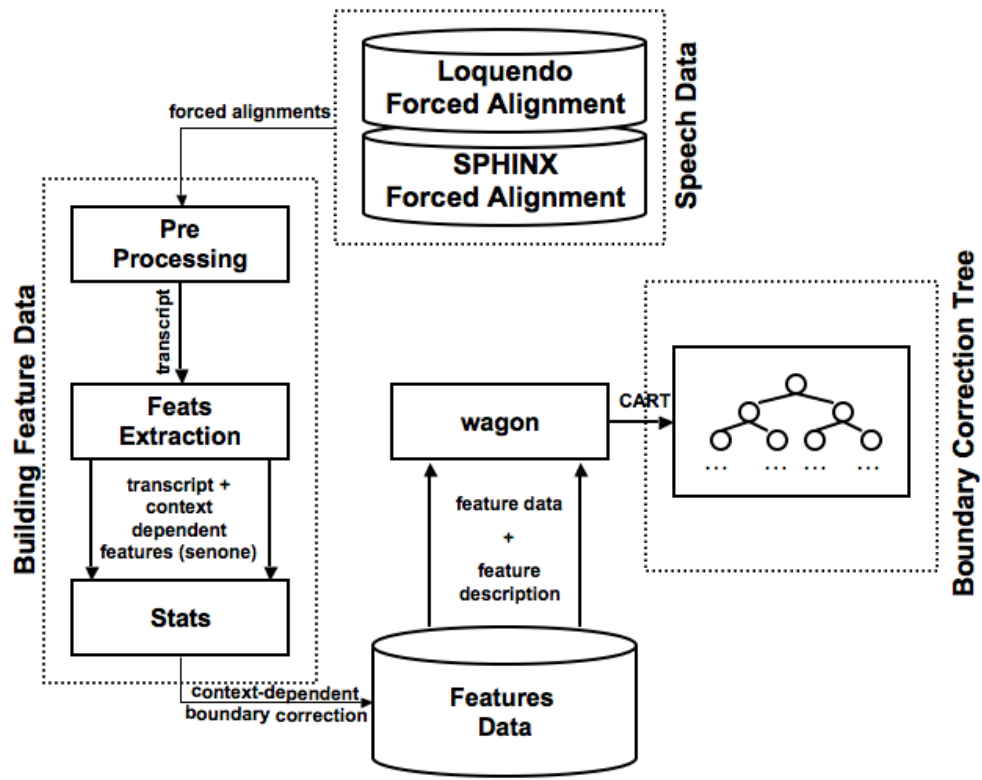

Figure 2: Graphical representation of the steps required to build a CART. From the training set a duplex forced alignment is extracted for each utterance. Both alignments are preprocessed in order to cope with the recognition errors and to achieve a unique, reliable transcription of the units. Then, from this transcription context-dependent information are extracted to build the relative senone. Each senone is then labeled with the systematic error computed by our error function. The software wagon gets as input the feature data (dataset) and the feature description (record description) and builds the regression tree.

## 6. RESULTS

Our model was trained "off-line" with respect to the construction of the prototype voice, using only a part of the utterances of our corpus and relating the automatic segmentation with the reference transcription. We refer to this part of the corpus as the *training set* (Table 1).

We evaluated the model from a statistical point of view in order to identify a measurement of the reduction of the error between the automatic speech segmentation of unknown utterances and their respective reference alignments, which are supposed to identify the optimal solution. We refer to the set of unknown utterances as the *test set* (Table 1).

| CARTree | Total error | Mean error | DEVSTD |
|---|---|---|---|
| **SPHINX-2** | 464.15 | 0.024 | 0.123 |
| | 69.03 | 0.028 | 0.185 |
| **cor.S5.tree** | 384.59 | 0.020 | 0.122 |
| | 61.44 | 0.025 | 0.185 |
| **cor.S10.tree** | 384.01 | 0.020 | 0.122 |
| | 62.31 | 0.025 | 0.186 |
| **cor.S25.tree** | 406.66 | 0.021 | 0.122 |
| | 61.39 | 0.025 | 0.185 |
| **cor.S100.tree** | 409.24 | 0.021 | 0.122 |
| | 62.41 | 0.025 | 0.183 |

Table 3: Summary of the experimental results. The first raw of each alignment refers to the values related to the training set, whilst the second row refers the values related to the test set. All the above values are in seconds.

Table 3 summarizes the experimental results collected by applying the regression models shown in Table 2. For each model's outcome alignment, the total difference in time, the mean error and the standard deviation with respect of the reference alignment are shown. The first row of each alignment refers the values related to the training set, whilst the second row refers to the values related to the test set.

For the aim of this paper, we only discuss the results relative to our best regression model called "cor.S25.tree" (for further details see (Zito, 2010)). The regression model applies a correction to each marker into the sphinx transcription following the context-dependent features of the relative phoneme building a new, more accurate, transcription.

Table 3 shows that we obtained a reduction of around 12% of the total error and confirms that the context-dependent units' boundary correction has a positive influence on the forced alignment.

In order to show the effects of the boundary correction on the context-dependent units, we present the test set as being composed of triphones in which each phone is characterized as a vowel (V), a consonant (C) or a pause (-). We use this representation to explain the context situation of the central phone of the triphone. Figure 3 shows the number of occurrences per each triphone in the test set.

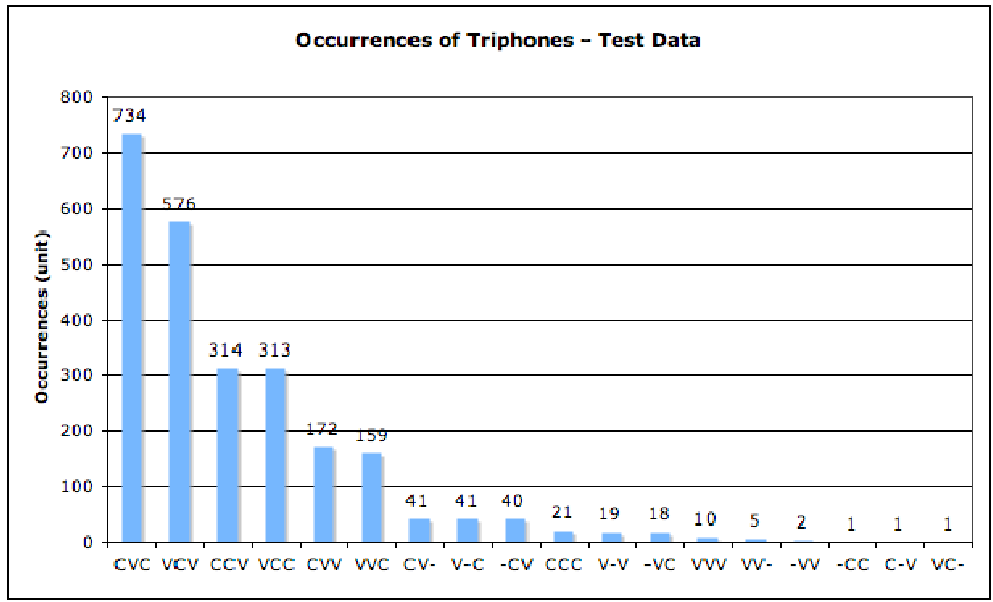

Figure 3: The chart shows the number of occurrences per each triphone in the test set. For a clearer reading the uncommon triphones are zoomed in on the top-right corner.

Figure 4 shows clearly that the automatic segmentation is less accurate in non-common context situations and, in fact, the mean errors are sensitively larger (orange columns). For example, in very uncommon contextual situations for the Italian language, as the combination consonant-pause-vowel (C-V), SPHINX-2 is not able to predict the correct boundary (in this specific example the mean error is larger than 0.05 sec), due to the lack of examples during the training. The standard approach to solve this problem requires further acoustic material which providing a better coverage of those samples and re-computing one or more training sessions. On the other hand, our approach demonstrated flexibility when faced with such a shortage and to produce sensible improvements.

Furthermore, the Figure 4 shows that our initial assumption to utilize the sign of the mean errors as an index of the direction in which to apply the corrections is reliable. In fact, there are only three cases where the correction is applied in the incorrect direction, and all these cases are characterized by a very small mean error (~0.01s).

Although, the mean error for the most common contextual situations does not show sensible improvements, the normalized error presented in Figure 5 put in evidence that the application of our method does not produce a sensible worsening of the segmentation.

## 7. DISCUSSION

In this study, we presented an innovative technique for speaker adaptation in order to improve the accuracy of segmentation with application to unit-selection Text-To-Speech (TTS) systems based on context dependent information extrapolated with linguistic analysis techniques.

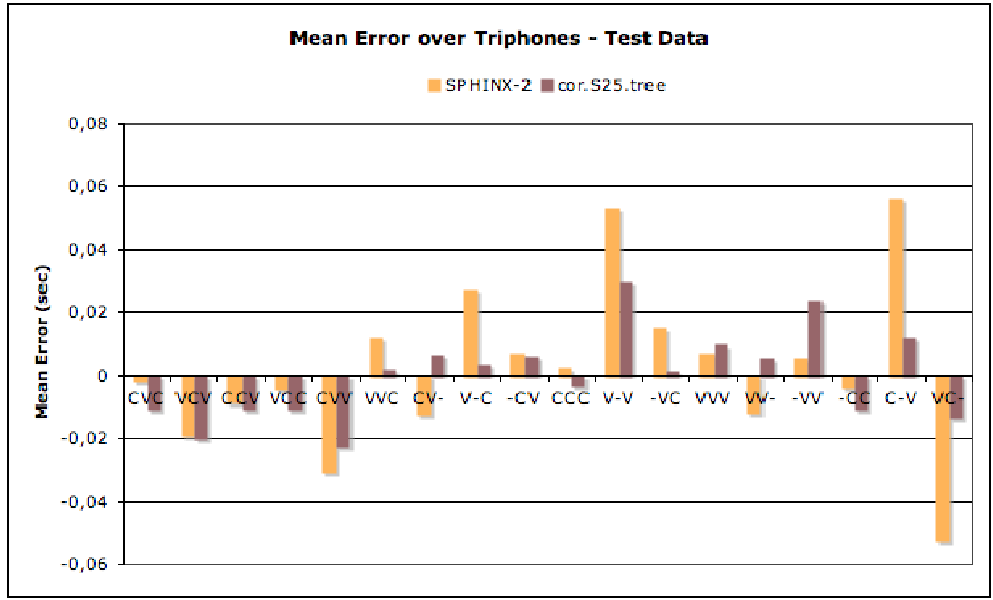

Figure 4: The chart shows the mean error made by SPHINX-2 before and after the correction using the "cor.S25.tree" model.

The results of this study showed that the context-dependent units boundary correction has a positive influence on the forced alignment, specifically when the misinterpretation of the phone is driven by acoustic properties linked to the speaker's phonetic characteristics.

The regression model proposed in this study attempts to maximize the performance evaluated with Root Mean Square Error (RMSE) and the correlation between the desired signal (which models the trend of the systematic error for each context-dependent phoneme) and the predicted signal. We obtained the best results with RMSE values equal to 0.040 and correlation of 0.444. We tested the model on around 3000 phonetic units and we obtained a reduction of around 12% of the total error with respect to the reference alignment, with positive performance also in the single context-dependent classes. However, the results showed that the performance of our system (in terms of percentage of error correction) is strongly limited to those phonemes for which the ASR computes an accurate segmentation. In fact, when the errors are shorter than one hundredth of second and with an arbitrary distribution, they cannot easily be modeled by our approach.

## 8. FUTURE WORK

We aim to reproduce the procedure on further corpora from the same speaker with the aim to release a new Italian voice with unit-selection synthesis for the FESTIVAL distribution.

The regression model will be compared with standard speaker-adaptation techniques in order to better understand its limitations and real advantages.

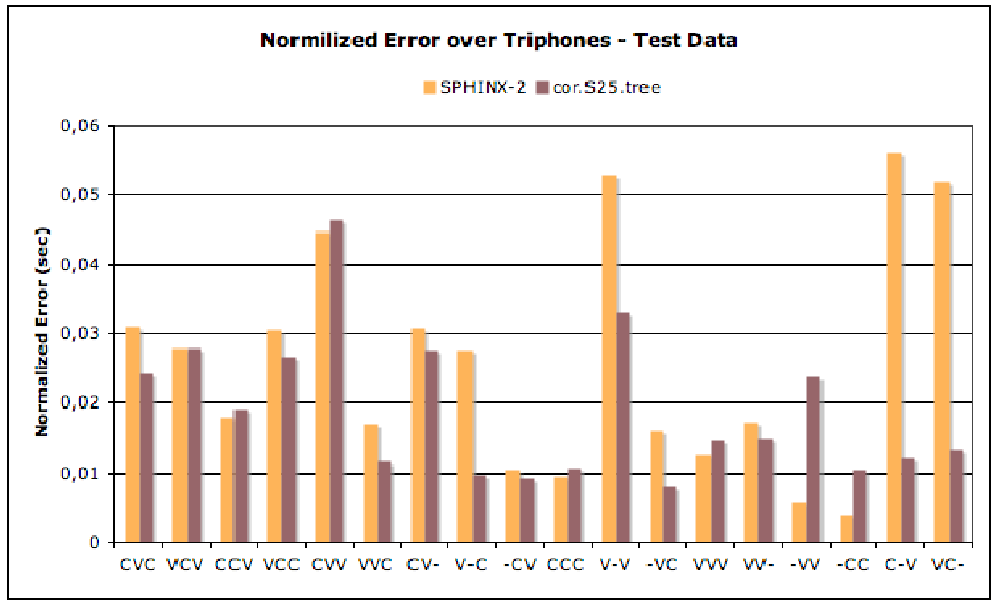

Figure 5: The chart shows the normalized mean error made by SPHINX-2 before and after the correction using the "cor.S25.tree" model.

## ACKNOWLEDGEMENTS

This work was supported by LOQUENDO, Italian company leader in speech technology, and partially by the EU FP7 "ALIZ-E" project (grant number 248116). Furthermore, the authors wish to acknowledge all the members of the Speech Unit of Hitachi Central Research Laboratory (Kokubunji, Tokyo, Japan) where this project has begun.